# Traffic incident analysis on urban arterials using ESE: A method for moderate length of time window

Zhen-zhen Yang [a,c], Liang Gao [a,*], Zi-you Gao [a], Ya-fu Sun [b,c], Sheng-min Guo [c,d]

[a] State Key Laboratory of Rail Traffic Control and Safety, MOE Key Laboratory for Urban Transportation Complex Systems Theory and Technology, Systems Science Institute, Center of Cooperative Innovation for Beijing Metropolitan Transportation, Beijing Jiaotong University, Beijing 100044, China

[b] Research and Development Center on Intelligent Transport Technology & Equipment, Ministry of Transport, Beijing 100191, China

[c] Beijing PalmGo Information and Technology Co., Ltd, Beijing 100191, China

[d] State Key Laboratory of Software Development Environment, Beihang University, Beijing 100191, China

**ABSTRACT**

Moderate length of time window can get the best accurate result in detecting the key incident time using extended spectral envelope. This paper presents a method to calculate the moderate length of time window. Two factors are mainly considered: (1) The significant vertical lines consist of negative elements of eigenvectors; (2) the least amount of interruption. The elements of eigenvectors are transformed into binary variable to eliminate the interruption of positive elements. Sine transform is introduced to highlight the significant vertical lines of negative elements. A novel Quality Index (QI) is proposed to measure the effect of different lengths of time window. Empirical studies on four real traffic incidents in Beijing verify the validity of this method.

**Key words**: traffic incident; extended spectral envelope; key incident time; length of time window; traffic phase transition

## 1. Introduction

Traffic incident usually generates bottleneck, which may arise traffic oscillations (Nie, 2010) and induce traffic phase transition (Daganzo et al., 1999) from free to congestion. The key incident time can be considered as the time when traffic oscillation or traffic phase transition happens. To relieve the problems caused by traffic incident, such as traffic congestion and environment pollution, many researches have been done to detect the key incident time. Existed methods can be mainly classified into two categories: statistical method (Gobol et al.,1987; Khattak et al.,1995; Garib et al.,1997; Wang, et al.,2008; Nam and Mannering, 2000; Jiyang et al., 2008; Zhan et al., 2011; Abdulla et al.,2011) and signal processing method (Muñoz and Daganzo, 2003; Sarvi et al., 2007; Li et al., 2011; 2012; 2014; Li et al., 2010; Zheng, 2011; Zhao et al., 2014).

---

* Corresponding author: Tel.: +86 (0)10 51687143.
  E-mail addresses: yangzhenzhen@bjtu.edu.cn (Z.Z. Yang), lianggao@bjtu.edu.cn (L. Gao), zygao@bjtu.edu.cn (Z.Y. Gao) , sunyafu@chinatransinfo.com(Y. F. Sun), guosm@nlsde.buaa.edu.cn (S.M. Guo).

Statistical methods include probabilistic distribution analysis (Gobol et al.,1987), time-sequential model (Khattak et al.,1995), linear regression analysis (Garib et al.,1997; Wang, et al.,2008), conditional probability analysis (Nam and Mannering, 2000), Bayesian decision tree (Jiyang et al., 2008), M5P tree algorithm (Zhan et al., 2011), fully parametric hazard-based duration models (Abdulla et al.,2011). Signal processing methods have oblique cumulative curves (Muñoz and Daganzo, 2003; Sarvi et al., 2007), describing-function approach (Li et al., 2011; 2012; 2014), frequency spectrum analysis approach (Li et al., 2010), wavelet transform (Zheng, 2011), extended spectral envelope (ESE) method (Zhao et al., 2014). Comparing to statistical methods, signal processing methods tend to dig the actual traffic flow characteristics and study the internal discipline further.

ESE is an extended method based on Spectral envelope (SE). SE method aims to emphasize salient oscillations by selecting certain transformations for the entire analysis period, which cannot be used to detect the start and end time of an oscillation, and not to investigate spatial-temporal evolution of traffic and transportation system (Zhao et al., 2014). Therefore, ESE is proposed to capture such dynamic properties. ESE method is the combination of long-term spectral envelope (LT-SE) and short-term spectral envelope (ST-SE) analysis. LT-SE analysis computes spectral envelope of all time windows at all locations. ST-SE analysis divides the study period into relatively short time windows and carries the SE analysis in each time window separately.

One of ESE method cores is to calculate the length of short time windows. Previous studies have proved that moderate length of time window can get the best accurate result. Zhao et al. (2014) found the window length by the difference between the peak of envelope and its closest valley. Empirical research has proven that this approach was not significant for traffic incident time detection (Yang, et al., 2015). So, this paper will propose a novel method to calculate the moderate length of time window.

The rest of this paper is organized as follows. Section 2 introduce the ESE method and the method of how to verdict the key incident time. Section 3 proposes a method for determining the moderate length of time window. Section 4 validates it by four traffic incidents in Beijing. Section 5 concludes the paper.

2. **ESE method**

In ESE method, study period $T$ is divided into short time windows. Define $L$ as the length of time window, $\eta$ as the overlap between the adjacent time windows, $\delta$ as the length of sliding window. $\varepsilon$ represents the ID of a time window. $N$ is the total number of time windows. i.e. $\varepsilon = 1,2,...,N$. The truncated time series at all location then form a truncated matrix.

$$V^\varepsilon = \begin{bmatrix} v_1^\varepsilon(1) & \cdots & v_1^\varepsilon(n) \\ \vdots & \ddots & \vdots \\ v_m^\varepsilon(1) & \cdots & v_m^\varepsilon(n) \end{bmatrix}$$

where $v_i^\varepsilon = [v_i^\varepsilon(1), v_i^\varepsilon(2), ..., v_i^\varepsilon(n)]$, $m$ is the number of all analyzed links, $n$ as the number of time series.

The spectral envelop of $V^\varepsilon$ at $\omega \in [-1/2, 1/2]$ is defined as

$$\lambda^\varepsilon(\omega) := \sup_{\beta^\varepsilon \neq 0} \left\{ \frac{P_S^\varepsilon(\omega, \beta^\varepsilon)}{C_S^\varepsilon(\beta^\varepsilon)} \right\} = \sup_{\beta^\varepsilon \neq 0} \left\{ \frac{(\beta^\varepsilon)^\Phi P_V^\varepsilon(\omega) \beta^\varepsilon}{(\beta^\varepsilon)^\Phi C_V^\varepsilon \beta^\varepsilon} \right\}$$

where $\beta^\varepsilon$ is a complex vector. $P_S^\varepsilon(\omega, \beta^\varepsilon)$ is the power concentrated at the frequency $\omega$, corresponding to the linear transformation defined by $\beta^\varepsilon$. $C_S^\varepsilon(\beta^\varepsilon)$ is the total power of the weighted time series. $P_V^\varepsilon(\omega)$ is the cross power spectral density matrix for $V^\varepsilon$ at the frequency $\omega$. The symbol "$\Phi$" denotes the conjugate transpose of the vector. $C_V^\varepsilon$ denotes the covariance matric associated with $V^\varepsilon$. The amplitude of $\lambda^\varepsilon(\omega)$ represents the largest portion of power that can be obtained from all transformed frequency $\omega$.

The spectral envelope equals the largest eigenvalue of matrix $D_V^\varepsilon \equiv (C_V^\varepsilon)^{-1/2} P_V^\varepsilon (C_V^\varepsilon)^{1/2}$, corresponding to an eigenvector

$$\varphi^\varepsilon = (C_V^\varepsilon)^{1/2} \beta^\varepsilon.$$

The key incident time can be found by detecting the sudden change of $\varphi^\varepsilon$. When most of the elements of $\varphi^\varepsilon$ change from positive to negative, a sudden phase transition of traffic flow happens at time window $\varepsilon$.

## 3. Method for moderate length of time window

The length of time window $L$ is a critical parameter for ESE method in detecting the key incident time. If $L$ is too short, too many negative elements of $\varphi^\varepsilon$ are created, and the key incident time points cannot be accurately identified. If $L$ is too long, the significant vertical lines will disappear. In this section, a method for calculating the moderate length of time window is proposed.

Two factors are mainly considered: (1) the significant vertical lines consist of negative elements of $\varphi^\varepsilon$, and (2) the least amount of interruption. To meet the two requirements, we first eliminate the effect of positive elements of $\varphi^\varepsilon$ by transforming the elements of $\varphi^\varepsilon$ into binary variable. Then sine transform is introduced to highlight the significant vertical lines of negative elements. A new

Quality Index (QI) is proposed to measure the effect of different lengths of time window.

$\varphi_{L,i}^{\varepsilon}$ is the element of $\varphi^{\varepsilon}$ with the length of time window $L$, where $L = 1,2,\ldots,n$. All eigenvectors with the length of time window $L$ form the following matrix.

$$\varphi_L = \begin{bmatrix} \varphi_{L,1}^1 & \cdots & \varphi_{L,1}^{\varepsilon} & \cdots & \varphi_{L,1}^N \\ \vdots & \cdots & \vdots & \cdots & \vdots \\ \varphi_{L,i}^1 & \cdots & \varphi_{L,i}^{\varepsilon} & \cdots & \varphi_{L,i}^N \\ \vdots & \cdots & \vdots & \cdots & \vdots \\ \varphi_{L,m}^1 & \cdots & \varphi_{L,m}^{\varepsilon} & \cdots & \varphi_{L,m}^N \end{bmatrix}$$

If $\varphi_{L,i}^{\varepsilon} \geq 0$, the phase of traffic flow does not transit; $\varphi_{L,i}^{\varepsilon} < 0$, the phase of traffic flow transit. To expediently describe the phase transition of traffic flow, the elements of $\varphi_L$ are transformed into binary variable.

$$b_{L,i}^{\varepsilon} = \begin{cases} 1, & if\ \varphi_{L,i}^{\varepsilon} < 0 \\ 0, & if\ \varphi_{L,i}^{\varepsilon} \geq 0 \end{cases}$$

All $b_{L,i}^{\varepsilon}$ form the following matrix.

$$b_L = \begin{bmatrix} b_{L,1}^1 & \cdots & b_{L,1}^{\varepsilon} & \cdots & b_{L,1}^N \\ \vdots & \cdots & \vdots & \cdots & \vdots \\ b_{L,i}^1 & \cdots & b_{L,i}^{\varepsilon} & \cdots & b_{L,i}^N \\ \vdots & \cdots & \vdots & \cdots & \vdots \\ b_{L,m}^1 & \cdots & b_{L,m}^{\varepsilon} & \cdots & b_{L,m}^N \end{bmatrix}$$

After transformed, all elements of $\varphi_L$ without traffic phase transition are zero, leaving the elements of $\varphi_L$ with traffic phase transition. In this way, we can specialize in the phase transition and extracting the appropriate length of time window.

The proportion of negative elements $\rho_L^{\varepsilon}$ is defined as

$$\rho_L^{\varepsilon} = \frac{1}{m} \sum_{i=1}^{m} b_{L,i}^{\varepsilon}$$

where $m$ is the number of analyzed links. All $\rho_L^{\varepsilon}$ form the following matrix

$$\rho_L = [\rho_L^1 \quad \cdots \quad \rho_L^{\varepsilon} \quad \cdots \quad \rho_L^N]$$

In the same way, we transfer $\rho_L^{\varepsilon}$ to binary variable.

$$f_L^\varepsilon = \begin{cases} 1, if\ \rho_L^\varepsilon > 0 \\ 0, if\ \rho_L^\varepsilon = 0 \end{cases}$$

Then, the number of vertical lines with the length of time window $L$ can be calculated by the following equation.

$$h_L = \sum_{\varepsilon=1}^{N} f_L^\varepsilon$$

All $h_L$ form the following matrix,

$$h = [h_1 \quad h_2 \quad \cdots \quad h_N].$$

To highlight the significant vertical lines of negative elements of $\varphi^\varepsilon$, the trigonometric function $\sin(\cdot)$ is introduced. The sine transform can convert the result symmetry to the horizontal axis, as illustrated in figure 1.

After the sine transform,

$$y = \begin{cases} sin(x - 0.5) < 0, x < 0.5 \\ sin(x - 0.5) = 0, x = 0.5 \\ sin(x - 0.5) > 0, x > 0.5 \end{cases}$$

When $x < 0.5$, $y < 0$; $x = 0.5$, $y = 0$; $x > 0.5$, $y > 0$.

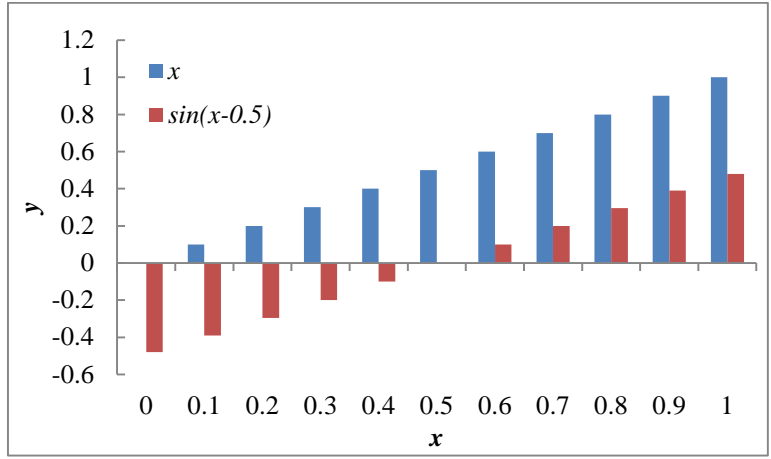

Figure 1. The sine transform

Define $\xi$ as the threshold to distinguish the high proportion of negative elements. Define $\alpha_L^\varepsilon$ as transformed results. Then

$$\alpha_L^\varepsilon = \begin{cases} sin(\rho_L^\varepsilon - \xi), if\ \rho_L^\varepsilon > 0 \\ 0, \qquad\qquad if\ \rho_L^\varepsilon = 0 \end{cases}$$

$\alpha_L$ is the average value of all $\alpha_L^\varepsilon$, then

$$\alpha_L = \frac{1}{h_L} \sum_{i=1}^{\varepsilon} \alpha_L^\varepsilon$$

All $\alpha_L$ form the following matrix,

$$\alpha = [\alpha_1 \quad \alpha_2 \quad \cdots \quad \alpha_n].$$

A new Quality Index (QI) is proposed to measure the effect of different lengths of time window.

$$QI = [QI_1 \quad \cdots \quad QI_L \quad \cdots \quad QI_n]$$

where

$$QI_L = \frac{\alpha_L}{h_L}$$

The moderate length of time window $L_{moderate}$ can be calculated by

$$L_{moderate} = \left\{ L \middle| \max_{1 \leq L \leq n} QI_L \right\}$$

## 4. Empirical study

### 4.1. Data

Four incidents in Beijing are analyzed by ESE method in our empirical study. Their moderate lengths of time window are calculated by the method presented in Section 3. The data used in this empirical study were collected from operating taxis in Beijing. Their locations and directions of traffic are showed in figure 2. The study period of four incidents are showed in table 1. For each incident, data are aggregated every 5 minutes as a time unit. Incident #1, #2 and #4 happened on the 4th Ring Road. Incident #3 happened on the 2th Ring Road.

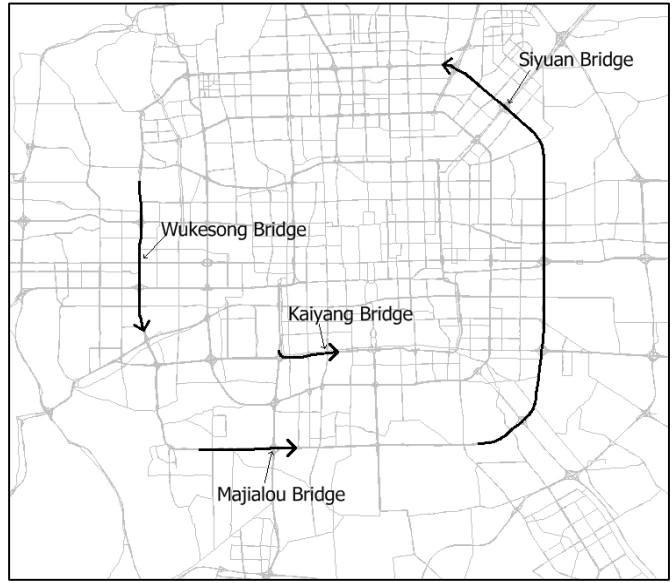

Figure 2. The empirical study area.

Table 1. The study period of four incidents

| Incident | Location | Study period | Date | Time Unit(s) |
|---|---|---|---|---|
| #1 | Siyuan Bridge | 05:00 to 24:00 | 2012-09-07 | 229 |
| #2 | Majialou Bridge | 17:00 to 18:50 | 2012-03-01 | 23 |
| #3 | Kaiyang Bridge | 16:35 to 18:30 | 2012-03-01 | 24 |
| #4 | Wukesong Bridge | 13:15 to 18:15 | 2012-03-01 | 61 |

**4.2. Results**

Set the threshold $\xi = 0.5$, the sliding window $\delta = 1$. Figure 3 is the results of four incidents. With the increase of the length of time window, the numbers of vertical lines $h$ are in a decreasing trend. After sine transform, the value of $\alpha$, whose proportion of negative elements $\rho_L^\varepsilon > \xi$, are above the zero-axis. The maximum QI for Incident #1, #2, #3, #4, are respectively 0.0357, 0.2955, 0.0552, 0.2173, with $L_{moderate} = 186, 19, 17, 24$. Figure 4 is the contour plot of the elements of the eigenvector. Figure 5 is the velocity at incident links. The detected incident time of Incident #1, #2, #3, #4, are respectively at 19, 3, 4, 27.

In figure 4(b), two vertical lines exist in incident #4. From figure 5(d), we can see that data between $18^{th}$ time unit and $21^{th}$ time unit disturbs the result of incident #4. Some other incident may influence the traffic. To better observe the result of incident #4, we select the data between $22^{nd}$ time unit and $61^{st}$ time unit to form a new data set, named incident #4-1.

Figure 6(a) is the results of incident #4-1. The maximum QI corresponds to $L = 33$. Figure 6 (b) is the contour plot of the elements of the eigenvector $\varphi^\varepsilon$ for each time window $\varepsilon$, where $\delta = 1$, $L = 33$. The vertical line appears at the $4^{th}$ time window. The incident time is at the $4^{th}$ time units in figure 6 (c), which is consistent with the result calculated by ESE in figure 6 (b).

The empirical study proves the efficiency of the method for moderate length of time window.

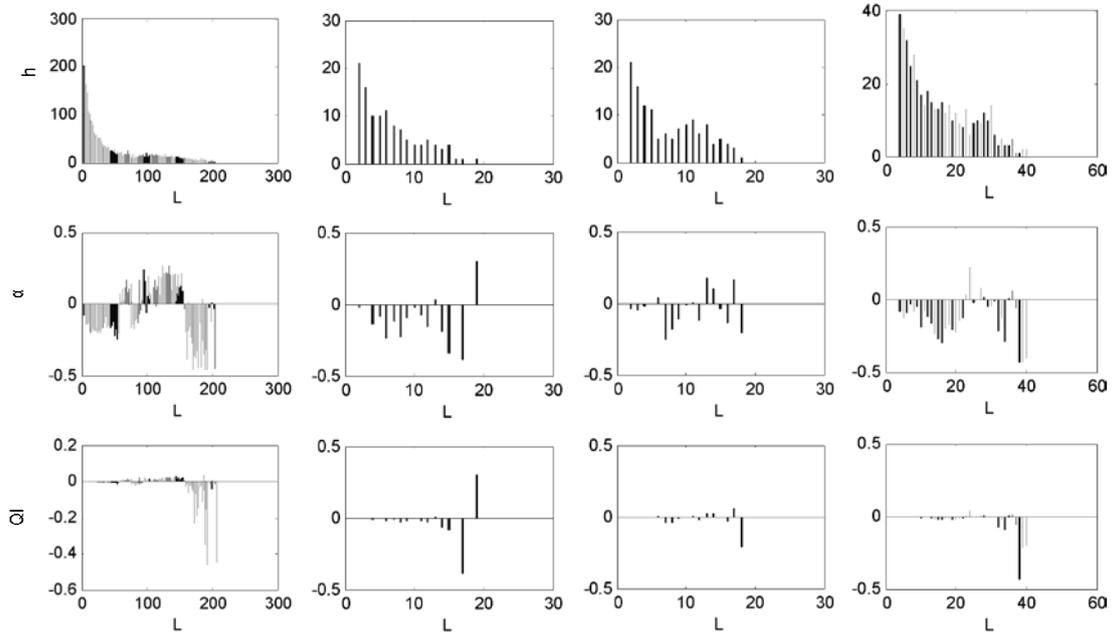

(a) Incident #1     (b) Incident #2     (c) Incident #3     (d) Incident #4

Figure 3. Results of four incidents. With the increase of time window length, the numbers of vertical lines are in the decreasing trend. When the proportion of negative elements $\rho_L^\varepsilon > \xi$, the value of $\alpha$ is above the zero-axis. The maximum QI for Incident #1, #2, #3, #4, are respectively 0.0357, 0.2955, 0.0552, 0.2173, with the moderate length of time window 186, 19, 17, 24.

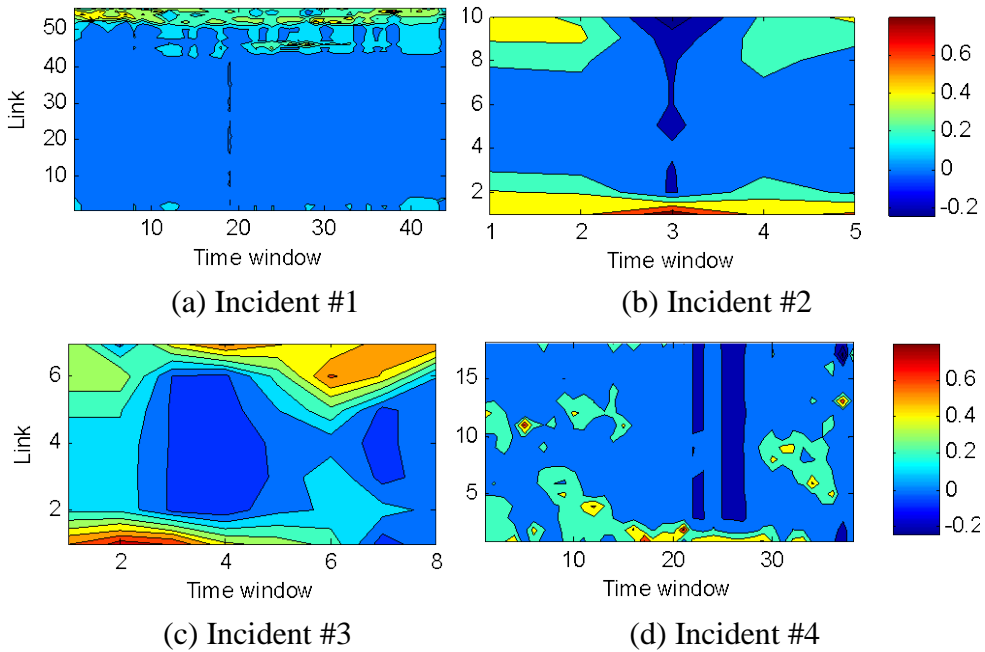

(a) Incident #1                    (b) Incident #2

(c) Incident #3                    (d) Incident #4

Figure 4. The contour plot of the elements of the eigenvector $\varphi^\varepsilon$ for each time window $\varepsilon$, where $\delta = 1$. (a) Incident #1, $L = 186$; (b) Incident #2, $L = 19$; (c)

Incident #3, $L = 17$; (d) Incident #4, $L = 24$.

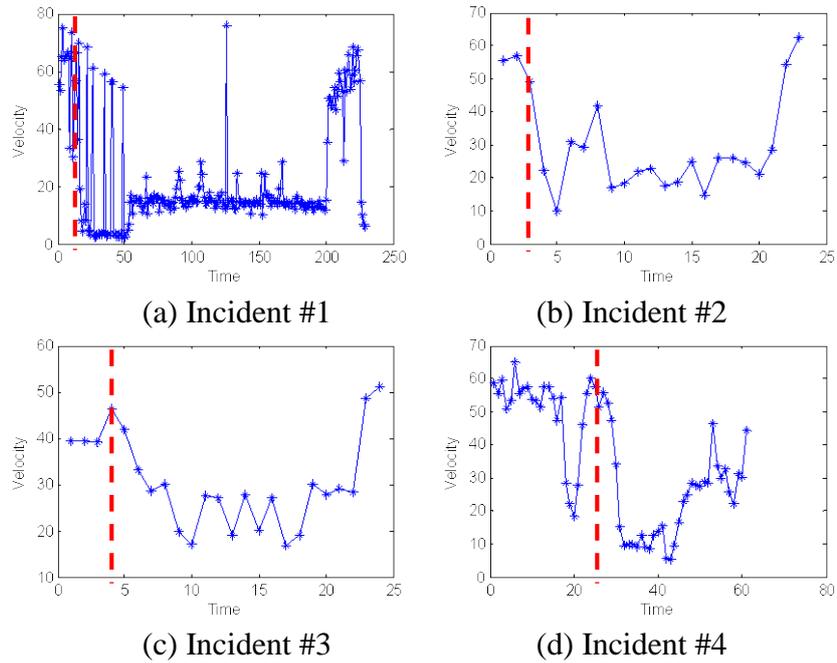

(a) Incident #1      (b) Incident #2

(c) Incident #3      (d) Incident #4

Figure 5. The velocity at incident links. The detected incident time of Incident #1, #2, #3, #4, are respectively at 19, 3, 4, 27.

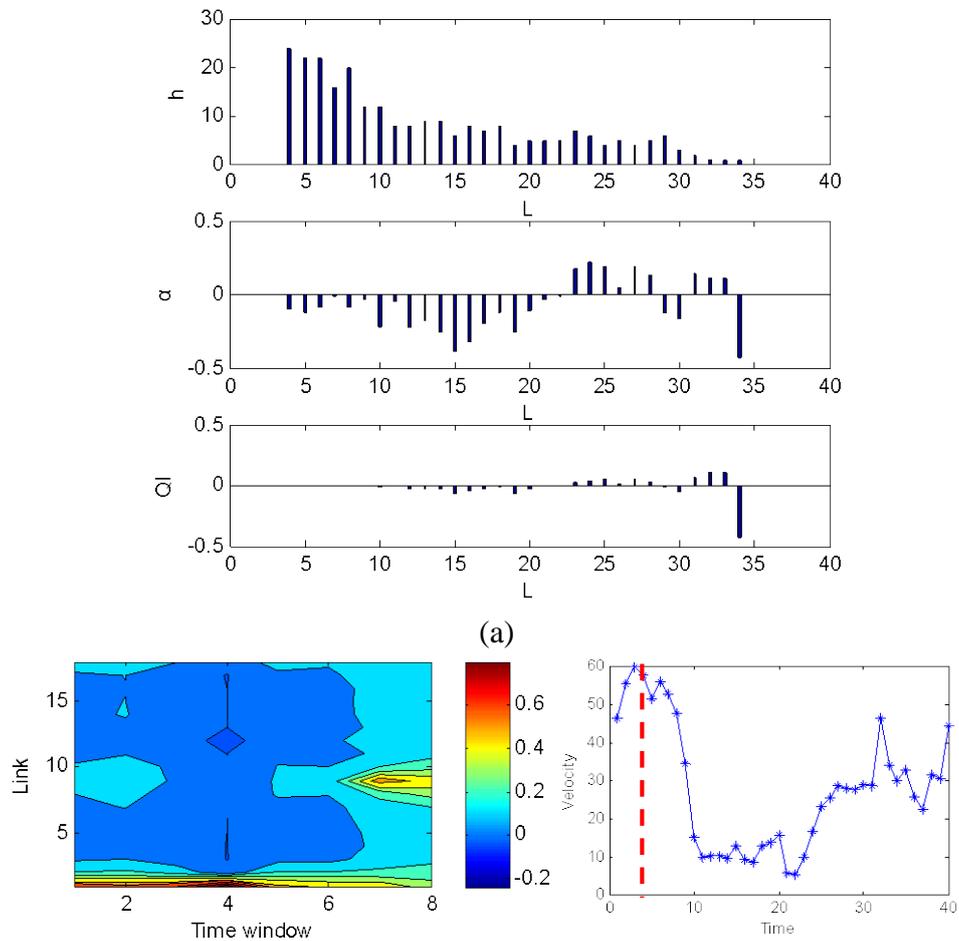

(a)

(b) (c)

Figure 6. Incident #4-1: (a) The results of $h$, $\alpha$, and $QI$ in the method for the time window length. The maximum $QI$ corresponds to $L = 33$. (b) The contour plot of the elements of the eigenvector $\varphi^\varepsilon$ for each time window $\varepsilon$, where $\delta = 1$, $L = 33$. (c) The velocity of incident link.

## 5. Conclusion

This paper presents a method for calculating the moderate length of time window when traffic incident analysis using extended spectral envelope (ESE). Four incidents are analyzed in the empirical study. The data used in this empirical study were collected from operating taxis in Beijing. Two factors are mainly considered: (1) The significant vertical lines consist of negative elements of eigenvectors; (2) The least amount of interruption as far as possible. The elements of eigenvectors are transformed into binary variable to eliminate the interruption of positive elements. Sine transform is introduced to highlight the significant vertical lines of negative elements. A new Quality Index (QI) is proposed to measure the effect of different lengths of time window. Empirical studies on four real traffic incidents in Beijing verify the validity of the proposed method.

All results indicating that the ESE method is providing a promising tool for the traffic incident analysis, and will foster the applications of traffic incident management system in ITS.

**Acknowledgement**

We thank the financial support from the Major State Basic Research Development Program of China (973 Program) No. 2012CB725400, the National High Technology Research and Development Program of China (863 Program) No. 2015AA124103, the National Natural Science Foundation of China (71101009, 71131001), the Fundamental Research Funds for the Central Universities No.2015JBM058. This work is partially supported by the State Key Laboratory of Rail Traffic Control and Safety (Contract No. RCS2014ZTY8), Beijing Jiaotong University.